\documentclass[twocolumn]{aastex6}
\usepackage{amsmath}
\usepackage{amssymb}
\usepackage{txfonts}
\fullcollaborationName{The Friends of AASTeX Collaboration}

\shorttitle{Satellitesimal formation via collisional dust growth}
\shortauthors{Shibaike et al.}

\begin{document}


\title{Satellitesimal Formation via Collisional Dust Growth in Steady Circumplanetary Disks}


\author{Yuhito Shibaike\altaffilmark{1}, Satoshi Okuzumi\altaffilmark{1}, Takanori Sasaki\altaffilmark{2} and Shigeru Ida\altaffilmark{3}}







\altaffiltext{1}{Department of Earth and Planetary Sciences, Tokyo Institute of Technology, Meguro-ku, Tokyo, 152-8551, Japan}
\altaffiltext{2}{Department of Astronomy, Kyoto University, Kitashirakawa-Oiwake-cho, Sakyo-ku, Kyoto, 606-8502, Japan}
\altaffiltext{3}{Earth-Life Science Institute, Tokyo Institute of Technology, Meguro-ku, Tokyo 152-8550, Japan}

\begin{abstract}
The icy satellites around Jupiter are considered to have formed in a circumplanetary disk. While previous models focused on the formation of satellites starting from satellitesimals, the question of how satellitesimals form from smaller dust particles has not been addressed so far. In this work, we study the possibility that satellitesimals form in situ in a circumplanetary disk. We calculate the radial distribution of the surface density and representative size of icy dust particles that grow by colliding with each other and drift toward the central planet in a steady circumplanetary disk with a continuous supply of gas and dust from the parent protoplanetary disk. The radial drift barrier is overcome if the ratio of the dust to gas accretion rates onto the circumplanetary disk, $\dot{M}_{\mathrm{d}}/\dot{M}_{\mathrm{g}}$, is high and the strength of turbulence, $\alpha$, is not too low. The collision velocity is lower than the critical velocity of fragmentation when $\alpha$ is low. Taken together, we find that the conditions for satellitesimal formation via dust coagulation are given by $\dot{M}_{\mathrm{d}}/\dot{M}_{\mathrm{g}}\ge1$ and $10^{-4}\le\alpha<10^{-2}$. The former condition is generally difficult to achieve, suggesting that the in-situ satellitesimal formation via particle sticking is viable only under an extreme condition. We also show that neither satellitesimal formation via the collisional growth of porous aggregates nor via streaming instability is viable as long as $\dot{M}_{\mathrm{d}}/\dot{M}_{\mathrm{g}}$ is low.
\end{abstract}




\section{Introduction} \label{introduction}

The gas giants in the solar system have large satellites whose orbits are on the same plane. These regular satellites are believed to have formed in gas disks around the planets, called circumplanetary disks, like protoplanetary disks that form planets around stars \citep{lun82}. Numerical simulations show that circumplanetary disks are by-products of gas accretion from the parent protoplanetary disk to the gas giants \citep[e.g.][]{lub99,dan02,tan12,fun16,szu16a}.

Satellite formation in circumplanetary disks has been studied extensively. \citet{lun82} proposed the minimum mass disk model where the circumplanetary disk is static and its solid mass is equal to the total mass of the satellites. \citet{mos03a}, \citet{mos03b}, \citet{est09}, and \citet{mos10} showed that the Jovian and Saturnian satellites can form in the minimum mass disks with a low-density outer portion, although the satellites would have to overcome rapid inward migration induced by the interaction with the massive disk \citep{mig16}. \citet{can02} and \citet{war10} proposed the gas-starved disk whose mass is regulated by viscous accretion onto the central planet and by a gradual supply of the gas from the parent protoplanetary disk. They showed that the temperature of the disk is sustained low enough to produce icy satellites. \citet{ali05} also developed a similar accretion disk model consistent with a formation model for Jupiter. \citet{can06} performed N-body simulations of satellite formation in the gas-starved disk, showing that the total mass of the satellites formed in the disk is $\sim10^{-4}$ of the central planet's mass as observed for the Jovian, Saturnian, and Uranian satellite systems. \citet{sas10} and \citet{ogi12} also studied the satellite formation in the gas-starved disk model using Monte Carlo and N-body simulations respectively, and successfully reproduced the number, masses, and orbits of the Galilean satellites. \citet{sas10} also reproduced the Saturnian system.
 
One significant problem of these satellite formation studies is that how satellitesimals form from dust particles are unaddressed. \citet{can06}, \citet{sas10}, and \citet{ogi12} assumed that the dust particles grow to satellitesimals as soon as they deliver from the protoplanetary to circumplanetary disks. However, it is already known from planetesimal formation studies that dust growth to kilometer-sized bodies in protoplanetary disks can be hindered by phenomena such as the radial drift and the collisional fragmentation of intermediate-sized particles \citep{whi72,ada76,wei77}. Therefore, it is easy to imagine that satellitesimal formation in circumplanetary disks could suffer from similar difficulties.

In this work, we aim to answer the question of whether dust particles can grow to satellitesimals by their direct collisional growth in circumplanetary disks. We employ a simple one-dimensional model in which we calculate the radial distribution of the surface density and typical size of dust particles in a steadily accreting circumplanetary disk. We also consider only icy dust particles and do not consider rocky particles. Although we assume perfect sticking upon collision, fragmentation occurs if the collision velocity is higher than a few $\rm{m~s^{-1}}$ when the aggregates are mainly composed of silicate particles \citep[e.g.][]{blu08,wad09}. The majority of the Galilean satellites are indeed icy satellites: Europa is $\sim10\%$ and Ganymede and Callisto are $\sim50\%$ ice by mass \citep{soh02}. Our simple treatment allows us to explore a large parameter space. The goal of this work is to derive the conditions under which satellitesimal formation via direct coagulation of dust particles is viable.

We note that it has also been discussed that planetesimals can be captured by gas drag from the circumplanetary disks \citep{fuj13,tan14,dan15,sue16,sue17}. \citet{sue16} and \citet{sue17} examined the captures and subsequent orbital evolutions of planetesimals. They showed that the capture hypothesis could roughly reproduce the initial radial distribution of planetesimals (i.e. satellitesimals) assumed in the satellite formation model by \citet{can06}, \citet{sas10}, and \citet{ogi12}. However, if there is a gas gap around the circumplanetary disk, the positive pressure gradient outside of the gap removes low eccentricity planetesimals from the feeding zone of the planet and the capture rate decreases \citep{kob12,fuj13,tan14,sue17}.

The plan of this paper is as follows. In Section \ref{methods}, we describe our model for the circumplanetary gas disk and explain how we treat the collisional growth and radial drift of the dust particles in the disk. In Section \ref{results}, we show the results of our calculations and derive the conditions for successful satellitesimal formation via dust coagulation. In Section \ref{discussions}, we discuss the feasibility of the conditions and consider other probabilities of  satellitesimal formation. Finally, we conclude this paper in Section \ref{conclusions}.

\section{Methods} \label{methods}
\subsection{Circumplanetary Disk Model} \label{diskmodel}
We model the structure of the circumplanetary disk following \citet{fuj14}. Although some numerical simulations suggested the possibility that gas near the midplane spiral outward \citep{tan12,fun16,szu16a}, here we assume that the circumplanetary disk is a viscous accretion disk with a continuous supply of material from the protoplanetary disk. The diffusion equation for the gas surface density $\Sigma_{\rm g}$ of the circumplanetary disk is then given by
\begin{equation}
\dfrac{\partial \Sigma_{\mathrm{g}}}{\partial t}=\dfrac{1}{r}\dfrac{\partial}{\partial r}\left[3r^{1/2}\dfrac{\partial}{\partial r}\left(r^{1/2}\nu\Sigma_{\mathrm{g}}\right)\right]+f,
\label{diffusion}
\end{equation}
where $r$ is the distance from the central planet, $f$ is the mass flux of the gas inflow from the protoplanetary to circumplanetary disks, and $\nu$ is the kinematic viscosity. We employ the standard $\alpha$ prescription \citep{sha73} and express the viscosity as $\nu=\alpha c_{\mathrm{s}}H_{\mathrm{g}}$, where $c_{\rm s}$ is the isothermal sound speed and $H_{\rm g}$ is the gas scale height. The sound speed is related to the temperature as $c_{\mathrm{s}}=\sqrt{k_{\mathrm{B}}T/m_{\mathrm{g}}}$ with $k_{\mathrm{B}}$ the Boltzmann constant and $m_{\mathrm{g}}=3.9\times10^{-24}~\mathrm{g}$ the mean molecular mass. The gas scale height is given by $H_{\mathrm{g}}=c_{\mathrm{s}}/\Omega_{\mathrm{K}}$, where $\Omega_{\mathrm{K}}=\sqrt{GM_{\mathrm{cp}}/r^{3}}$ is the Kepler frequency, and $G$ and $M_{\mathrm{cp}}$ are the gravitational constant and the central planet mass, respectively. Unless otherwise noted, we assume $M_{\mathrm{cp}}$ to be the Jupiter mass $M_{\mathrm{J}}=1.89\times10^{30} ~\mathrm{g}$. Based on the results of the three-dimensional hydrodynamical simulation by \citet{tan12}, \citet{fuj14} modeled $f$ as $f \propto r^{-1}$ for $r<r_{\rm b}$ and $f = 0$ for $r> r_{\rm b}$, where $r_{\rm b}$ is the radius of the region where the gas falls in. With this scaling for $f$, the steady-state solution of Equation~\eqref{diffusion} can be analytically obtained as \citep[see Equations~23 and 25 of][]{fuj14}
\begin{equation}
\Sigma_{\mathrm{g}}=\dfrac{\dot{M}_{\rm g}}{2\pi r_{\rm b}} \frac{r^{3/2}}{\nu}\left(-\dfrac{2}{9}r^{-1/2}+\dfrac{2}{3}r_{\mathrm{b}}r^{-3/2}\right),
\label{sigmag}
\end{equation}
where $\dot{M}_{\mathrm{g}}$ is the mass accretion rate of the infall gas. The simulation by \citet{tan12} shows that $r_{\rm b} \approx 20R_{\rm J}$ for the planet of $M_{\rm cp} = 0.4M_{\mathrm{J}}$ \citep[see also][]{fuj14}. Assuming that $r_{\rm b}$ scales with the Hill radius of the central planet, we use $r_{\rm b} = 27R_{\rm J}$ for our $1M_{\mathrm{J}}$-mass planet.

We assume that the circumplanetary disk is viscously heated. Then, the gas temperature at the midplane is given by \citet{nak94},
\begin{equation}
T=\left( \dfrac{9}{8 \sigma_{\rm SB}}\nu\Sigma_{\mathrm{g}}\Omega_{\mathrm{K}}^{2}\right)^{1/4} g(\tau),
\label{T4}
\end{equation}
where $\sigma_{\mathrm{SB}}$ is the Stefan-Boltzmann constant and
\begin{equation}
g(\tau)=\left(\dfrac{3}{8}\tau+\dfrac{1}{4.8\tau}\right)^{1/4}
\label{g}
\end{equation}
is a function of the Rosseland mean optical depth $\tau$. In principle, $\tau$ depends on the size distribution of the smallest dust particles, which cannot be predicted with simple dust evolution models as employed in this study. Lacking good knowledge about $\tau$, we opt to set $g\approx1$. Since $g\ga1$ in general, the assumption $g\approx1$ yields a minimum estimate for the disk temperature. The temperature can be up to three times higher than assumed here if the optical depth ranges between $10^{-2}\lesssim\tau\lesssim10^{2}$. However, this uncertainty has little effects on the main results of this work because the dependence of our results on $T$ is weak (see Section \ref{gasdust}).

According to the simulation by \citet{tan12}, the mass flux $\dot{M}_{\rm g}$  scales as $\dot{M}_{\rm g}\approx0.2 \Sigma_{\rm PPD}r_{\rm H}^2\Omega_{\rm PPD}$, where $\Sigma_{\rm PPD}$ and $\Omega_{\rm PPD}$ are the gas surface density and orbital period of the parent protoplanetary disk  in the vicinity of the planet, respectively, and $r_{\rm H}$ is the planet' Hill radius  \citep[see Figure 14 of][]{tan12}. At Jupiter's orbit, the gas surface density is $143~\mathrm{g\ cm^{-2}}$ according to the minimum-mass solar nebula model of \citet{hay81}. For this value of $\Sigma_{\rm PPD}$, the accretion rate onto Jupiter-sized planet would be $\dot{M}_{\mathrm{g}}\approx 200~M_{\rm J}~{\rm Myr}^{-1}$.

\begin{figure*}
\epsscale{1.15}
\plotone{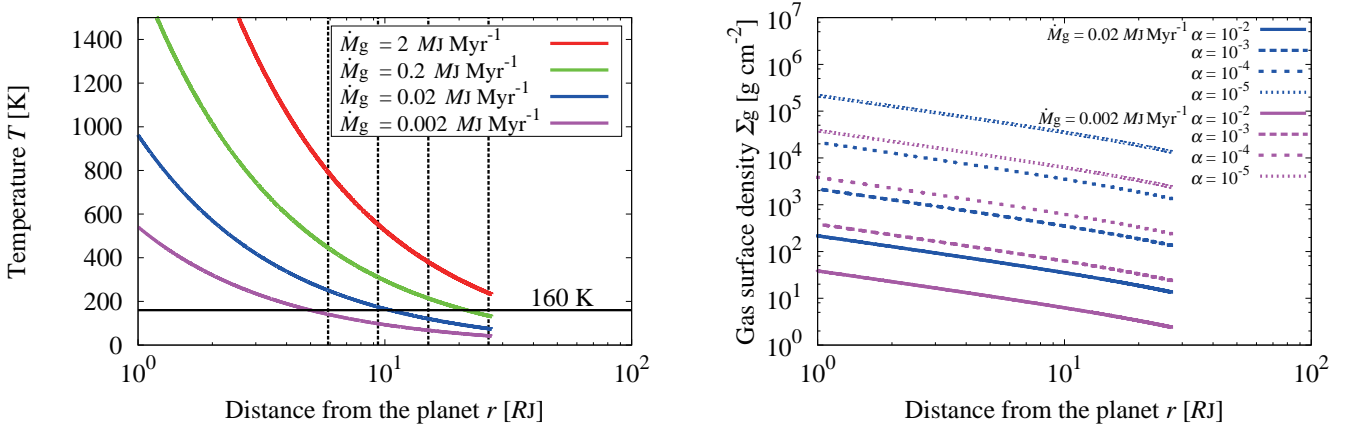}
\caption{Gas surface density and the midplane temperature of the circumplanetary disk. The red, blue, green, and purple curves in the left panel represent the temperatures, where $\dot{M}_{\mathrm{g}}=2, 0.2, 0.02$, and $0.002~M_{\rm J}~{\rm Myr}^{-1}$, respectively. The horizontal black line is the sublimation temperature of icy aggregates. The vertical black dashed lines represent the current orbits of the Galilean satellites. The blue and purple curves in the left panel represent the gas surface density, where $\dot{M}_{\mathrm{g}}=0.02$ and $0.002~M_{\rm J}~{\rm Myr}^{-1}$, respectively. The strength of the turbulence is $\alpha=10^{-5}$, $10^{-4}$, $10^{-3}$, and $10^{-2}$. \label{fig:Gas}}
\end{figure*}

However, a strong constraint on $\dot{M}_{\rm g}$ can be obtained from the temperature of the circumplanetary disk. In the left panel of Figure \ref{fig:Gas}, we plot the midplane temperature of our modeled circumplanetary disk as a function of the distance from the central planets for three cases $\dot{M}_{\mathrm{g}}=2, 0.2, 0,02$, and $0.002 ~M_{\rm J}~{\rm Myr}^{-1}$. For $\dot{M}_{\mathrm{g}}=2$ and $0.2~M_{\rm J}~{\rm Myr}^{-1}$, the temperatures are higher than the sublimation temperature of ice, which is about 160 K. Such hot environments are unsuitable for the formation of icy regular satellites around Jupiter and Saturn. Therefore, we only consider $\dot{M}_{\mathrm{g}}\le0.02~M_{\rm J}~{\rm Myr}^{-1}$ in this work. We plot the gas surface density of the disk for the two cases in the right panel of Figure \ref{fig:Gas}.

\subsection{Dust Growth and Radial Drift} \label{dustgrowth}
We treat the coagulation and radial drift of dust particles in circumplanetary disks. In particular, we focus on steady state where the radial distribution of the size and surface density of the particles is constant over time. We approximate the size distribution of the particles at each distance from the central planet by a narrow distribution peaked at  mass $m_{\rm d}$. With this approximation, the integro-differential equation governing the evolution of particles in a disk can be rewritten into a simple differential equation for $m_{\rm d}$. We also approximate the radial distribution of dust inflow to the circumplanetary disk by a narrow peak lying at $r=r_{\mathrm{b}}$, the outer edge of the infall region. With this approximation, the problem of obtaining the radial distribution of $\Sigma_{\rm d}$ and $m_{\rm d}$ in steady state reduces to a simple boundary-value problem. We discuss the validity of these assumptions in Section \ref{validity}.

In addition to the above approximations, we for the moment assume that the collision velocity between the particles is so low that their fragmentation is negligible. The issue of the fragmentation barrier will be separately treated in Section \ref{turbulence}. We also assume that the initial particle radius (i.e., the particle radius at $r = r_{\rm b}$) is $0.1\ \mathrm{mm}$. This is the maximum size of the particles which can diffuse into the gas gap against the outward drift motion caused by the positive gas pressure gradient at the gap's outer edge \citep{zhu12}. We focus on steady state where the radial distribution of the mass and surface density of the particles is independent of time.

In steady state, and in the absence of fragmentation, the mass $m_{\rm d}$ of the radially drifting particles is determined as a function of $r$ by \citep[Equation (5) of][]{sat16}
\begin{equation}
v_{\mathrm{r}}\dfrac{d m_{\mathrm{d}}}{d r}=\dfrac{2\sqrt{\pi}R_{\mathrm{d}}^{2}\Delta v_{\mathrm{dd}}}{H_{\mathrm{d}}}\Sigma_{\mathrm{d}},
\label{growth}
\end{equation}
where $R_{\mathrm{d}}$, $\Delta v_{\mathrm{dd}}$, $v_{\mathrm{r}}$, $H_{\mathrm{d}}$ and $\Sigma_{\mathrm{d}}$ are the radius, collision velocity, drift speed, scale height, and surface mass density of the particles, respectively. The particle mass is related to the particle radius by $m_{\mathrm{d}}=(4\pi/3) R_{\mathrm{d}}^{3}\rho_{\mathrm{int}}$, where $\rho_{\mathrm{int}}$ is the internal density of the particles. We fix $\rho_{\mathrm{int}}=1.4~\mathrm{g~cm^{-3}}$ but we discuss the effects of changing it in Section \ref{internal}. The inward accretion rate of the particles is
\begin{equation}
\dot{M_{\mathrm{d}}}=-2\pi rv_{r}\Sigma_{\mathrm{d}},
\label{continuous}
\end{equation}
which is, in steady state, constant over $r$ and is equal to the infall rate set at $r = r_{\rm d}$ as a boundary condition. We numerically integrate Equation (\ref{growth}) with Equation (\ref{continuous}) from $r = r_{\rm b}$ toward smaller $r$.

The radial drift velocity of a dust particle is determined by its stopping time, $t_{\mathrm{stop}}$. In this study, we express the stopping time in terms of the Stokes number defined by $\mathrm{St}=\Omega_{\mathrm{K}}t_{\mathrm{stop}}$ In dense circumplanetary disks, one can safely assume that particles of $R_{\rm d} >$ 0.1 mm are much larger than the mean free path of gas molecules so that the flow around the particles can be regarded as continuous fluid. Then, the Stokes number can be expressed as
\begin{equation}
\mathrm{St}=\dfrac{8}{3C_{\mathrm{D}}}\dfrac{\rho_{\mathrm{int}}R_{\mathrm{d}}}{\rho_{\mathrm{g}}\Delta v_{\mathrm{dg}}}\Omega_{\mathrm{K}},
\label{stokes_newton}
\end{equation}
where $\Delta v_{\mathrm{dg}}$ is the relative velocity between the dust particles and the gas, and $C_{\mathrm{D}}$ is a dimensionless coefficient that depends on the particle Reynolds number, $\mathrm{Re_{p}}$. According to \citet{per11}, the coefficient can be written as
\begin{equation}
C_{\mathrm{D}}=\dfrac{24}{\mathrm{Re_{p}}}(1+0.27\mathrm{Re_{p}})^{0.43}+0.47(1-\exp(-0.04\mathrm{Re}_{\rm p}^{0.38})).
\label{CD}
\end{equation}
The particle Reynolds number  is given by
\begin{equation}
\mathrm{Re_{p}}=\dfrac{4R_\mathrm{{d}}\Delta v_{\mathrm{dg}}}{v_{\mathrm{th}}\lambda_{\mathrm{mfp}}},
\label{Rep}
\end{equation}
where $v_{\mathrm{th}}=\sqrt{8/\pi}c_{\mathrm{s}}$ is the thermal velocity and $\lambda_{\mathrm{mfp}}=m_{\mathrm{g}}/(\sigma_{\mathrm{mol}}\rho_{\mathrm{g}})$ is the mean free pass of the gas 
with $\sigma_{\mathrm{mol}}=2\times10^{-15}\mathrm{cm}^{2}$ the collisional cross section of the gas molecules and $\rho_{\mathrm{g}}=\Sigma_{\mathrm{g}}/(\sqrt{2\pi}H_{\mathrm{g}})$ the gas density at the midplane.

The scale height of the particles can be derived analytically from the balance of their vertical sedimentation and diffusion \citep{you07},
\begin{equation}
H_{\mathrm{d}}=H_{\mathrm{g}}\left(1+\dfrac{\mathrm{St}}{\alpha}\dfrac{1+2\mathrm{St}}{1+\mathrm{St}}\right)^{-1/2}.
\label{Hd}
\end{equation}
The radial drift velocity of the dust particles is \citep{whi72,ada76,wei77}
\begin{equation}
v_{\mathrm{r}}=-2\dfrac{\mathrm{St}}{\mathrm{St}^{2}+1}\eta v_{\mathrm{k}},
\label{vr}
\end{equation}
where $v_{\mathrm{k}}=r\Omega_{\mathrm{k}}$ is the Kepler velocity and
\begin{equation}
\eta=-\dfrac{1}{2}\left(\dfrac{H_{\mathrm{g}}}{r}\right)^{2}\dfrac{\partial \ln{\rho_{\mathrm{g}}c_{\mathrm{s}}^{2}}}{\partial \ln{r}}
\label{eta}
\end{equation}
is the ratio of the pressure gradient force to the gravity of the central planet.

The relative velocity between the dust particles (i.e. collision velocity) is the root sum square
\begin{equation}
\Delta v_{\mathrm{dd}}=\sqrt{\Delta v_{\mathrm{B}}^{2}+\Delta v_{\mathrm{r}}^{2}+\Delta v_{\mathrm{\phi}}^{2}+\Delta v_{\mathrm{z}}^{2}+\Delta v_{\mathrm{t}}^{2}},
\label{vdd}
\end{equation}
where $\Delta v_{\mathrm{B}}$, $\Delta v_{\mathrm{r}}$, $\Delta v_{\mathrm{\phi}}$, $\Delta v_{\mathrm{z}}$, and $\Delta v_{\mathrm{t}}$ are the relative velocities induced by Brownian motion, the radial drift, azimuthal drift, vertical sedimentation, and turbulence \citep{oku12}. For collisions between equal-sized particles, the Brownian-motion-induced velocity can be written as $\Delta v_{\mathrm B}=\sqrt{16k_{\mathrm{B}}T/(\pi m_{\mathrm{d}})}$. The relative velocity induced by the radial drift is $\Delta v_{\mathrm{r}}=|v_{\mathrm{r}}(\mathrm{St}_{1})-v_{\mathrm{r}}(\mathrm{St}_{2})|$, where $\mathrm{St}_{1}$ and $\mathrm{St}_{2}$ are the Stokes numbers of the two particles. We assume $\mathrm{St}_{2}=0.5\mathrm{St}_{1}$ \citep[see Section 2.4 in][]{sat16} and $v_{\mathrm{r}}$ is given by Equation (\ref{vr}). The relative velocity induced by the azimuthal drift is $\Delta v_{\mathrm{\phi}}=|v_{\mathrm{\phi}}(\mathrm{St}_{1})-v_{\mathrm{\phi}}(\mathrm{St}_{2})|$, where $v_{\mathrm{\phi}}=-\eta v_{\mathrm{K}}/(1+\mathrm{St}^{2})$. The relative velocity induced by the vertical motion is $\Delta v_{\mathrm{z}}=|v_{\mathrm{z}}(\mathrm{St}_{1})-v_{\mathrm{z}}(\mathrm{St}_{2})|$, where $v_{\mathrm{z}}=-\Omega_{\mathrm{K}}\mathrm{St}H_{\mathrm{d}}/(1+\mathrm{St})$. For the relative velocity induced by the turbulence, we use the analytic formula derived from \citet{orm07}. The formula has three limiting expressions:
\begin{equation}
\Delta v_{\mathrm{t}}=
\begin{cases}
     \sqrt{\alpha}c_{\mathrm{s}}{\rm Re}_{\rm t}^{1/4}|\mathrm{St}_{1}-\mathrm{St}_{2}|, &  \mathrm{St}_{1}\ll {\rm Re}_{\rm t}^{-1/2}, \\
     \sqrt{3\alpha}c_{\mathrm{s}}\mathrm{St}_{1}^{1/2}, & {\rm Re}_{\rm t}^{-1/2} \ll \mathrm{St}_{1}\ll 1, \\
     \sqrt{\alpha}c_{\mathrm{s}}\left(\dfrac{1}{1+\mathrm{St}_{1}}+\dfrac{1}{1+\mathrm{St}_{2}}\right)^{1/2}, & 1\ll \mathrm{St}_{1}.
\end{cases}
\label{vt}
\end{equation}
Here, $\mathrm{Re_{t}}=\nu/\nu_{\mathrm{mol}}$ is the turbulence Reynolds number, where $\nu_{\mathrm{mol}}= v_{\mathrm{th}}\lambda_{\mathrm{mfp}}/2$ is the molecular viscosity. We obtain the relative velocity between the solid materials and the gas, $\Delta v_{\mathrm{dg}}$, by setting $\mathrm{St}_{1}=\mathrm{St}$ and 
$\mathrm{St}_{2}\rightarrow0$ in the above expressions for the relative velocities.

\subsection{Parameter Choice} \label{parameter}
Table \ref{tab:parameterchoice} summarizes the parameter range explored in this study. The gas infall rate $\dot{M}_{\mathrm{g}}$ onto the circumplanetary disk is taken to be either $0.02~M_{\rm J}~{\rm Myr}^{-1}$ or $0.002~M_{\rm J}~{\rm Myr}^{-1}$. As mentioned in Section~\ref{diskmodel}, we do not consider a higher value of $\dot{M}_{\mathrm{g}}$ since the disk would become too hot for icy satellites to form. In reality, a giant planet carves a gap around its orbit. For example, the hydrodynamical simulations by \citet{kan15} show that the gas surface density inside the gap is depleted by a factor of more than 100 compared to outside the gap. This means that realistic values of $\dot{M}_{\mathrm{g}}$ should be less than $2~M_{\rm J}~{\rm Myr}^{-1}$, i.e., less than 1\% of the accretion rate without a gap. The infall rate that adopted in the gas-starved disk model of \citet{can02} is about $0.2~M_{\rm J}~{\rm Myr}^{-1}$. The value of $\dot{M}_{\mathrm{g}}$ whose temperature is suitable for icy satellite formation is lower than these estimated values. Although we do not consider the decrease of the gas inflow in detail, the final phase of planetary formation should be suitable for satellite formation (see also Section \ref{feasibility}).

The ratio $\dot{M}_{\mathrm{d}}/\dot{M}_{\mathrm{g}}$ of the dust inflow rate to the gas inflow rate  is chosen between 0.001--1. If we assume that the inflow has the solar composition and the dust particles are strongly coupled with the gas, the ratio should be $0.01$. However, the gap structure of the gas around Jupiter dams the dust particles drifted from the outer region of the protoplanetary disk toward the Sun and makes their similar gap structure. On the other hand, the strong gas gradient may trigger a hydrodynamic instability and disturb the gas inflow. This disturbance should enhance the radial diffusion of small dust particles and make them nearer to the central planet \citep{zhu12,tan14}.

The strength $\alpha$ of turbulence is varied from $10^{-5}$ to  $10^{-2}$. We note that mechanisms that could drive turbulence in the circumplanetary disk is highly uncertain. The magneto-rotational instability, a viable mechanism driving turbulence in ionized accretion disks, could operate on the surface of circumplanetary disks \citep{tur14}, but might not produce fully developed turbulence \citep{fuj14}. Therefore, we cannot rule out that $\alpha$ of circumplanetary disks falls below $\alpha$. However, as we discuss in the following section, too weak turbulence would make it difficult for satellitesimals to form within a realistic range of $\dot{M}_{\mathrm{d}}/\dot{M}_{\mathrm{g}}$.

\begin{table}
\begin{center}
\caption{Parameter choice}
\begin{tabular}{lll} \hline
Quantity & Description & Value \\ \hline \hline
$\dot{M}_{\mathrm{g}}$ & Gas infall rate & $0.02$, $0.002~M_{\rm J}~{\rm Myr}^{-1}$ \\
$\dot{M}_{\mathrm{d}}/\dot{M}_{\mathrm{g}}$ & Dust-to-gas infall rate ratio & $1$, $0.1$, $0.01$, $0.001$ \\ 
$\alpha$ & Turbulence parameter & $10^{-5}$, $10^{-4}$, $10^{-3}$, $10^{-2}$ \\ \hline
\end{tabular}
\label{tab:parameterchoice}
\end{center}
\end{table}

\section{Results} \label{results}
In this section, we present the results of our dust growth calculations and explore the conditions under which satellitesimals can form througu dust coagulation in circumplanetary disks.

\subsection{Fiducial Calculations} \label{evolution}
\begin{figure}
\epsscale{1.15}
\plotone{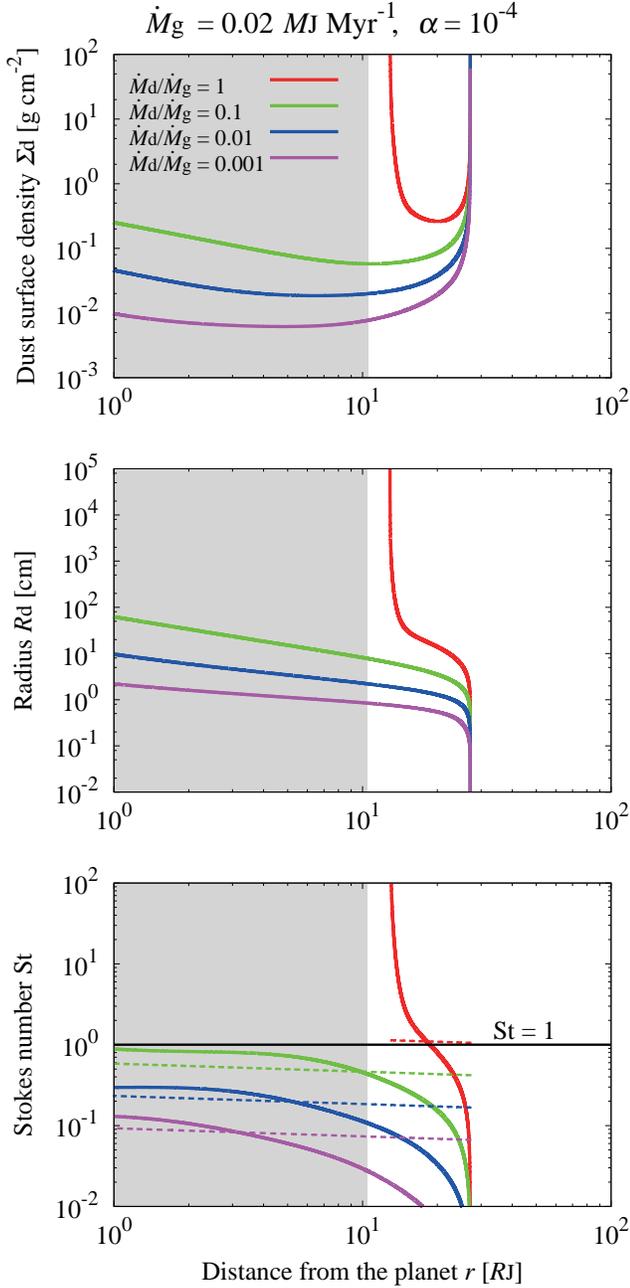}
\caption{Steady-state profiles of the surface density $\Sigma_{\rm d}$ (top panel) and radius $R_{\rm d}$ (middle panel) of dust particles that grow and drift in the circumplanetary disk of $\dot{M}_{\mathrm{g}}=0.02~M_{\rm J}~{\rm Myr}^{-1}$ and $\alpha=10^{-4}$. The bottom panel shows the Stokes number ${\rm St}$ of the particles. The red, green, blue, and purple curves correspond to $\dot{M}_{\mathrm{d}}/\dot{M}_{\mathrm{g}}=1,0.1,0.01$, and $0.001$, respectively. The dashed lines in the bottom panel show the prediction from the analytic estimate given by Equation (\ref{St}). Shaded in gray is the region interior to the snow line, which lies at $r=10 R_{\rm J}$. \label{fig:fiducial}}
\end{figure}

Figure~\ref{fig:fiducial} shows the results of our fiducial calculations that assume $\dot{M}_{\mathrm{g}}=0.02~M_{\rm J}~{\rm Myr}^{-1}$ and $\alpha=10^{-4}$. We assume that the snow line is where the midplane temperature is $160~{\rm K}$ and it is at $r=10 R_{\rm J}$ (see Figure (\ref{fig:Gas})). The top, middle, and bottom panels represent the surface density, radius, and Stokes number of the mass-dominating dust particles in the disk as a function of the distance $r$ from the central planet. Because the particles grow and move inward at the same time, the particle size increases with decreasing $r$. As already mentioned in the previous section, we assume that the dust particles are 0.1 mm in size when they are initially delivered from the protoplanetary disk to $r = r_{\mathrm{b}} (= 27R_{\mathrm{J}})$. However, the middle panel of \ref{fig:fiducial} suggests that the assumption about the initial size of the dust particles is not crucial because the particles immediately grow at $r \approx r_{\mathrm{b}}$. The change of $r_{\mathrm{b}}$ will not affect the steady-state profiles either because the profiles should gradually approach the approximated lines (see the dashed lines in the bottom panel of Figure~\ref{fig:fiducial} and Equation (\ref{St})).

As the particles grow, their Stokes number ${\rm St}$ and inward drift velocity $|v_{\rm r}|$ increases in accordance with Equations~\eqref{stokes_newton} and \eqref{vr}. We find that the radial drift becomes appreciable when their drift timescale $t_{\mathrm{drift}}=r/|v_{\mathrm{r}}|$ becomes shorter than 30 times the growth timescale $t_{\rm grow}=m_{\mathrm{d}}/(\mathrm{d}m_{\mathrm{d}}/\mathrm{d}t)$, in agreement with the situation for dust evolution in protoplanetary disks \citep{oku12,tsu17}. For $\dot{M}_{\mathrm{d}}/\dot{M}_{\mathrm{g}} = 1$, we find that the particles stop drifting and grow to kilometer-sized satellitesimals at $r \sim 10~R_{\rm J}$. The drift stalls because the drift speed (normalized by $\eta v_{\mathrm{K}}$) decreases with increasing size as long as ${\rm St} > 1$. Therefore, they have to overcome this barrier of the $\mathrm{St}=1$ for growing to the satellitesimals. After they achieve $\mathrm{St}=1$, the drift speed becomes slower and they get jammed. The jam makes the collisional rate higher, so that the  collisional growth speeds up. We also found that this condition $\mathrm{St}>1$ is consistent with $t_{\mathrm{grow}}<t_{\mathrm{drift}}$ when $\mathrm{St}\sim1$. Note that we did not consider the possibility that the dust surface density near the snowline increases because of sublimation or recondensation \citep[e.g.][]{sai11,ros13,ida16a,ida16b,sch17}.

\subsection{Effects of the Dust and Gas Inflow Mass Fluxes} \label{gasdust}
The amount of the gas and dust that flow to the circumplanetary disk can be changed by the conditions of the central planet, the protoplanetary disk, and the circumplanetary disk. We investigate the effects of changing the gas and dust inflow mass fluxes. Figure \ref{fig:fiducial} shows that the the dust surface density increases with the dust-to-gas inflow mass flux ratio $\dot{M}_{\mathrm{d}}/\dot{M}_{\mathrm{g}}$. The radius and Stokes number of the dust particles also have the same features. The particles can grow to satellitesimals only when $\dot{M}_{\mathrm{d}}/\dot{M}_{\mathrm{g}}=1$. This can be understood by using the approximate analytical expression for the Stokes number (dotted lines in the figure). When $\alpha=10^{-4}$ and $\dot{M}_{\mathrm{g}}=0.02~M_{\rm J}~{\rm Myr}^{-1}$, the gas surface density is so large that $\mathrm{Re_{p}}\ll1$. For example, we found that the particle Reynolds number $\mathrm{Re_{p}}$ is about $10^{3}$ when $\dot{M}_{\mathrm{d}}/\dot{M}_{\mathrm{g}}=0.1-1$ at $r\sim10~R_{\rm J}$. In this case, the dimensionless coefficient $C_{\mathrm{D}}$ can be approximated as a constant, $C_{\mathrm{D}}\approx0.5$ (Newton's friction law, see Equation (\ref{CD})). The dust--dust and dust--gas relative velocities can also be approximated as $\Delta v_{\mathrm{dd}}\approx(1/2)v_{\mathrm{r}}$ and $\Delta v_{\mathrm{dg}}\approx v_{\mathrm{r}}$ because the turbulence is so week that $\Delta v_{\mathrm{t}}$ is much smaller than $\Delta v_{\mathrm{r}}$ (see Figure \ref{fig:velocities}). The approximated Stokes number can then be described as,
\begin{equation}
\begin{split}
\mathrm{St}\approx&1.2\left(\dfrac{\dot{M}_{\mathrm{d}}/\dot{M}_{\mathrm{g}}}{1}\right)^{2/5}\left(\dfrac{\alpha}{10^{-4}}\right)^{1/5} \\
&\times\left(\dfrac{T}{160~\mathrm{K}}\right)^{-2/5}\left(\dfrac{M_{\mathrm{cp}}}{1~M_{\mathrm{J}}}\right)^{2/5}\left(\dfrac{r}{10~R_{\mathrm{J}}}\right)^{-2/5},
\label{St}
\end{split}
\end{equation}
for small $r$. Equation (\ref{St}) is derived by substituting Equation (\ref{continuous}) into Equation (\ref{growth}) and integrating it. Here, the gas surface density and the midplane temperature have been approximated as $\Sigma_{\mathrm{g}}\approx\dot{M}_{\mathrm{g}}\Omega_{\mathrm{K}}/(3\pi\alpha c_{\mathrm{s}}^{2})$ and $T\approx(3GM_{\mathrm{cp}}\dot{M}_{\mathrm{g}}/(8\pi\sigma_{\mathrm{SB}}r^{3}))^{1/4}$. The scale height and radial drift velocity of the dust particles have also been approximated as $H_{\mathrm{d}}\approx H_{\mathrm{g}}(\alpha/\mathrm{St})^{1/2}$ and $v_{\mathrm{r}}\approx-2\mathrm{St}\eta v_{\mathrm{K}}$. Equation (\ref{St}) shows that the Stokes number is proportional to $(\dot{M}_{\rm{d}}/\dot{M}_{\rm{g}})^{2/5}$ and reaches unity when $\dot{M}_{\rm{d}}/\dot{M}_{\rm{g}}=1$ and $r=10~R_{\rm{J}}$. Once $\mathrm{St}$ exceeds unity, the radial drift velocity starts to decrease with increasing particle size, and hence the particles grow to satellitesimals. Neglecting the weak $\dot{M}_{\rm{g}}$ dependence of $T$ ($T\propto\dot{M}_{\mathrm{g}}^{1/4}$), $\Sigma_{\mathrm{g}}$ is proportional to $\dot{M}_{\mathrm{g}}$. When $\Sigma_{\mathrm{g}}$ is high, the gas drag force that the dust particles receive is strong and $\rm{St}$ is small. The collision rate of the dust particles becomes high when $\dot{M}_{\mathrm{d}}$ (i.e. $\Sigma_{\mathrm{d}}$) is large. The high collision rate promotes satellitesimal formation. Even if the Newton's friction law does not apply, the Stokes number is proportional to $(\dot{M}_{\rm{d}}/\dot{M}_{\rm{g}})^{2/3}$ or $(\dot{M}_{\rm{d}}/\dot{M}_{\rm{g}})^{6/11}$ and the trend that the Stokes number is an increasing function does not change (see Equations (\ref{Stg}) and (\ref{Stp}) in Appendix \ref{app}).

\begin{figure}
\epsscale{1.15}
\plotone{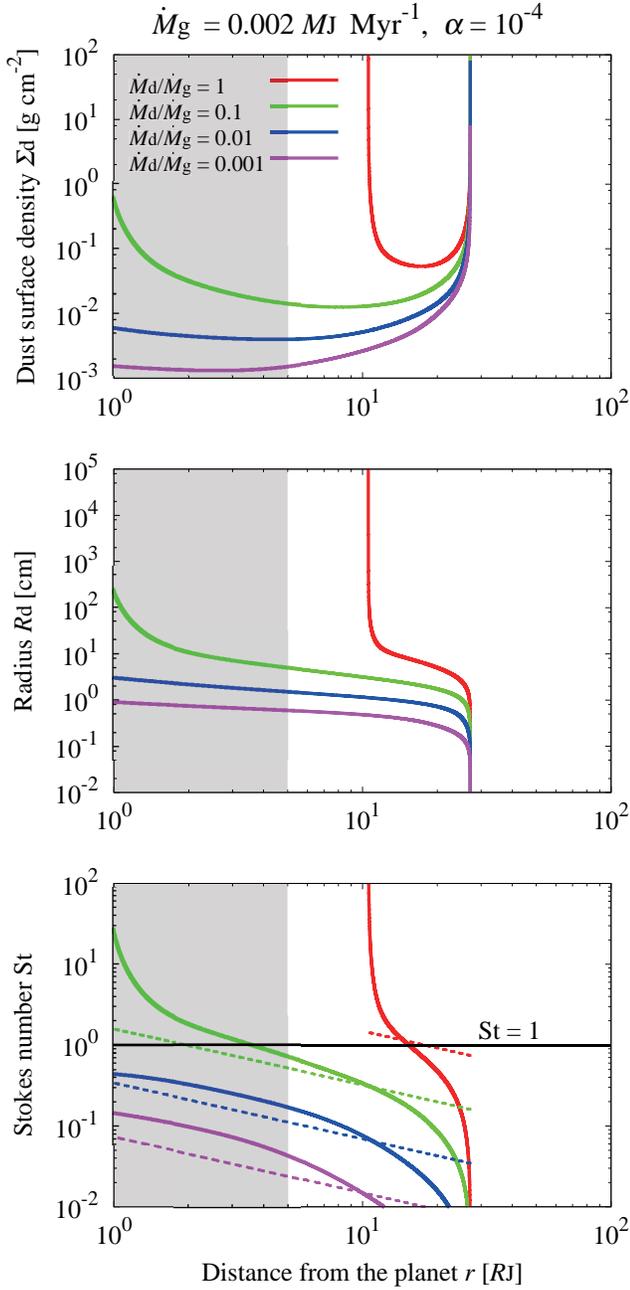}
\caption{Same as Figure~\ref{fig:fiducial}, but for $\dot{M}_{\rm g}=0.002~M_{\rm J}~{\rm Myr}^{-1}$ and $\alpha = 10^{-4}$. The dashed lines in the bottom panel show the prediction from the analytic estimate given by Equation (\ref{Stg}). Shaded in gray is the region interior to the snow line, which lies at $r=5 R_{\rm J}$. \label{fig:mdotg}}
\end{figure}

The key parameter of the dust evolution is not the pure gas inflow mass flux but the ratio of the dust and gas inflow mass fluxes. Figure \ref{fig:mdotg} represents the distributions of the dust surface density, the dust radius, and the Stokes number of the representative dust particles for $\dot{M}_{\mathrm{g}}=0.002~M_{\rm J}~{\rm Myr}^{-1}$ and $\alpha=10^{-4}$. In this case, the snow line lies at $r=5 R_{\rm J}$ (see Figure (\ref{fig:Gas})). The profiles of the surface density and radius of dust particles are lower than those for $\dot{M}_{\mathrm{g}}=0.02~M_{\rm J}~{\rm Myr}^{-1}$ and $\alpha=10^{-4}$ (Figure \ref{fig:fiducial}). The radial profiles of $\rm{St}$ are steeper than those in the fiducial case (we derive Equation (\ref{Stg}), an analytic equation of $\rm{St}$ for $\dot{M}_{\rm{g}}=0.002~M_{\rm J}~{\rm Myr}^{-1}$, in Appendix \ref{app}). Nevertheless, we find that dust particles grow beyond $\mathrm{St}=1$ only when $\dot{M}_{\rm{d}}/\dot{M}_{\mathrm{g}}=1$. For fixed $\dot{M}_{\mathrm{d}}/\dot{M}_{\mathrm{g}}$,  $\dot{M}_{\rm{g}}$ dependence of $\rm{St}$ is indeed weak. When $\dot{M}_{\rm{g}}=0.02~M_{\rm J}~{\rm Myr}^{-1}$, only $T$ depends on $\dot{M}_{\mathrm{g}}$ ($T\propto\dot{M}_{\mathrm{g}}^{1/4}$) so that $\mathrm{St}\propto T^{-2/5}\propto\dot{M}_{\mathrm{g}}^{-1/10}$  (Equation (\ref{St})). When $\dot{M}_{\rm{g}}=0.002~M_{\rm J}~{\rm Myr}^{-1}$ or $\alpha=10^{-2}$, $\mathrm{St}\propto T^{-1}\times\dot{M}_{\mathrm{g}}^{4/9}\propto\dot{M}_{\mathrm{g}}^{7/36}$ or $\mathrm{St}\propto T^{-1}\times\dot{M}_{\mathrm{g}}^{4/11}\propto\dot{M}_{\mathrm{g}}^{5/44}$ (Equations (\ref{Stg}) and (\ref{Stp})).

\subsection{Effects of the Strength of Turbulence}
\label{turbulence}
The strength of turbulence in the circumplanetary disk is also a key parameter of the dust evolution. Figure \ref{fig:mdotg} represents the profiles of the dust particles in the case with $\dot{M}_{\mathrm{g}}=0.02~M_{\rm J}~{\rm Myr}^{-1}$, and $\alpha=10^{-5}$ (upper panel) and $\alpha=10^{-2}$ (lower panel). In the case of $\alpha=10^{-5}$, the Stokes number of drifting particles is on average lower than those in the fiducial case with $\alpha=10^{-4}$. Even if $\dot{M}_{\rm{d}}/\dot{M}_{\rm{g}}=1$, the dust particles cannot grow to satellitesimals outside of the snow line at $r=10~R_{\rm J}$. In the case of $\alpha=10^{-2}$, the Stokes number is slightly higher than in the fiducial case. Equations (\ref{St}) and (\ref{Stp}) show that the Stokes number is actually proportional to $\alpha^{1/5}$ or $\alpha^{1/11}$ when $\dot{M}_{\mathrm{g}}=0.02~M_{\rm J}~{\rm Myr}^{-1}$ (see also Equation (\ref{Stg}) for $\dot{M}_{\mathrm{g}}=0.002~M_{\rm J}~{\rm Myr}^{-1}$). However, the stokes number is not high enough for the particles to overcome the radial drift outside the snow line unless $\dot{M}_{\rm{d}}/\dot{M}_{\rm{g}}\geq1$.

Even if $\dot{M}_{\rm{d}}/\dot{M}_{\rm{g}}=1$, satellitesimals would not form via direct dust growth because the relative velocity between the dust particles would be too high to avoid collisional fragmentation. Collision simulations by \citet{wad09} argued that icy dust aggregates with monomers of $0.1~\mathrm{\mu m}$ fragment upon collision if the collision velocity is higher than $50~\rm{m~s^{-1}}$. Figure \ref{fig:velocities} represents the dust--dust relative velocities for different values of $\alpha$. When $\alpha=10^{-4}$, the relative velocity is determined by the radial drift speed and that induced by the turbulence is low. When $\alpha=10^{-2}$, the dust--dust relative velocity is determined by the turbulence because the relative velocity induced by the turbulence is proportional to $\sqrt{\alpha}$ (Equation(\ref{vt})) and it becomes 10 times higher than that for $\alpha=10^{-4}$. Figure \ref{fig:velocities} shows that the relative velocity exceeds $50~\rm{m~s^{-1}}$, indicating that collisional fragmentation would happen. Therefore, satellitesimal formation via direct dust coagulation is unlikely to occur in such strong turbulence. Note that experiments by \citet{gun15} showed that the fragmentation occurs with the collision speed of $\sim10~\mathrm{m~s^{-1}}$ for icy aggregates with monomers of $1~\mathrm{\mu m}$.

\begin{figure}
\epsscale{1.15}
\plotone{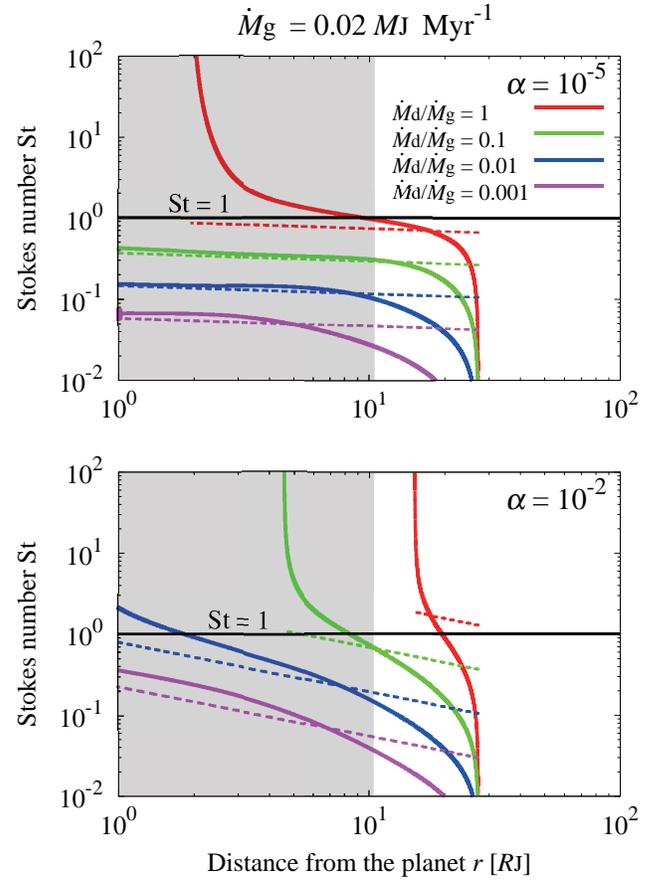}
\caption{Steady-state profiles of the Stokes number ${\rm St}$ of the particles when $\dot{M}_{\rm g} = 0.02~M_{\rm J}~{\rm Myr}^{-1}$, and $\alpha = 10^{-5}$ (upper panel) and $\alpha = 10^{-2}$ (lower panel). The dashed lines in the lower panel show the prediction from the analytic estimate given by Equation (\ref{Stp}). \label{fig:alpha}}
\end{figure}

\begin{figure*}
\epsscale{1.15}
\plotone{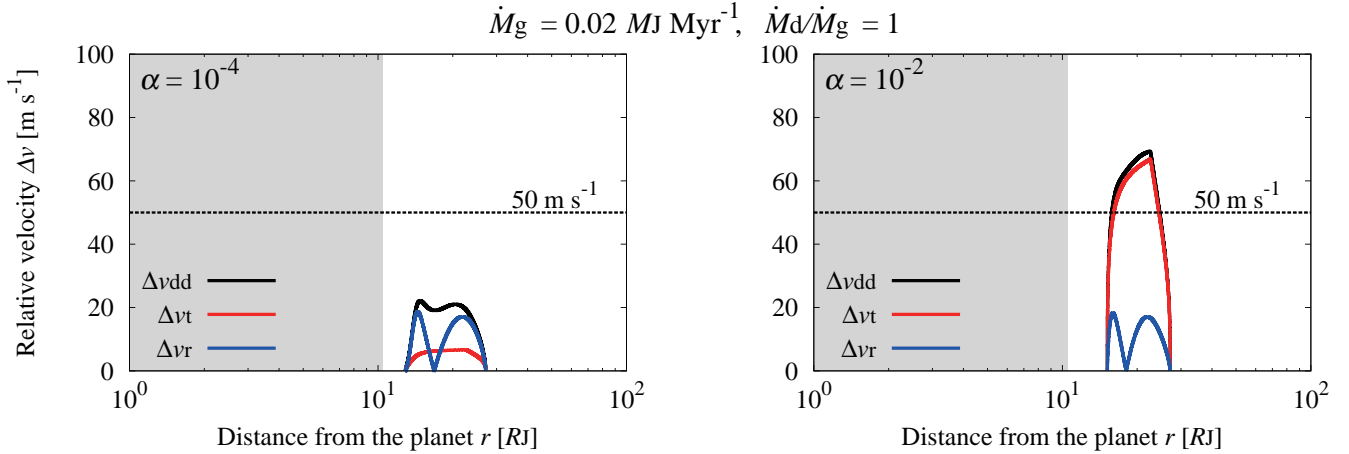}
\caption{Dust--dust relative velocities with different turbulence strength, $\alpha=10^{-4}$ (left panel) and $\alpha=10^{-2}$ (right panel). The other conditions are $\dot{M}_{\mathrm{d}}/\dot{M}_{\mathrm{g}}=1$ and $\dot{M}_{\mathrm{g}}=0.02~M_{\rm J}~{\rm Myr}^{-1}$ in both the panels. The black curves represent the dust--dust relative velocities (collision velocities). The red and blue curves represent the velocities induced by only the turbulence and their radial drift, respectively. The black dashed lines are the critical velocity of fragmentation. \label{fig:velocities}}
\end{figure*}

\subsection{Conditions for Satellitesimal Formation} \label{condition}
\begin{figure}
\epsscale{1.15}
\plotone{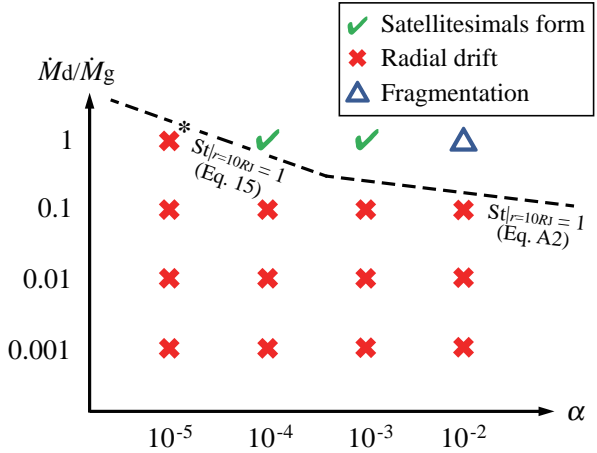}
\caption{Condition for satellitesimal formation when $\dot{M}_{\mathrm{g}}=0.02~M_{\rm J}~{\rm Myr}^{-1}$. The green ticks indicate that dust particles grow to satellitesimals outside of the snow line at $r=10~R_{\mathrm{J}}$. The red crosses indicate that the radial drift barrier inhibits dust growth to satellitesimals. The blue triangles indicate that dust particles grow to satellitesimals on the calculations but the collision velocity (dust--dust relative velocity) is faster than the critical velocity of fragmentation, $50~\mathrm{m~s^{-1}}$. The dashed lines show the condition $\mathrm{St}=1$ at $r=10~R_{\mathrm{J}}$ from Equations (\ref{St}) and (\ref{Stp}). When $\dot{M}_{\mathrm{g}}=0.002~M_{\rm J}~{\rm Myr}^{-1}$, the condition is the same except that the drift barrier is overcome outside the snow line if $\dot{M}_{\mathrm{d}}/\dot{M}_{\mathrm{g}}=1$ and $\alpha=10^{-5}$ (asterisk). \label{fig:condition}}
\end{figure}

The results presented in the previous subsections can be summarized in Figure \ref{fig:condition}. This figure represents the condition for satellitesimal formation when $\dot{M}_{\mathrm{g}}=0.02~M_{\rm J}~{\rm Myr}^{-1}$. The conditions are $\dot{M}_{\mathrm{d}}/\dot{M}_{\mathrm{g}}\ge1$ and $10^{-4}\le\alpha<10^{-2}$. The condition for breaking through the radial drift barrier is approximately given by $\dot{M}_{\mathrm{d}}/\dot{M}_{\mathrm{g}}>6\times10^{-3}\times\alpha^{-1/2}$ derived from the condition $\rm{St>1}$ at $r=10~R_{\rm J}$ (see Equation (\ref{St})). When the turbulence is strong ($\alpha\ga10^{-3}$), it is about $\dot{M}_{\rm{d}}/\dot{M}_{\rm{g}}>0.08~\alpha^{-1/6}$ (see Equation (\ref{Stp})). The dashed lines in Figure (\ref{fig:condition}) show the boundary of each condition. However, in the case of $\alpha=10^{-2}$, the aggregate collision velocity is too high to avoid collisional fragmentation. When $\dot{M}_{\rm{g}}=0.002~M_{\rm J}~{\rm Myr}^{-1}$, the drift barrier is overcome outside the snow line even if $\alpha=10^{-5}$ because the line is at $r=5~{R_{\rm J}}$ (asterisk in Figure (\ref{fig:condition})).

\section{Discussions} \label{discussions}
\subsection{Feasibility of the High Dust-to-Gas Inflow Mass Flux Ratio} \label{feasibility}
We found that the one of the conditions for satellitesimal formation is $\dot{M}_{\mathrm{d}}/\dot{M}_{\mathrm{g}}\ge1$. However, this condition may be difficult to achieve. First, the dust particles tend to settle down toward the midplane, the inflow gas from the high altitude is likely to dust-poor gas \citep{tan12}. This effect must depend on the conditions of the turbulence and the gas density of the region around the circumplanetary disk which the accretion gas comes from (Equations (\ref{St}) and (\ref{Hd})). Second, the dust supply may not be enough to achieve $\dot{M}_{\mathrm{d}}/\dot{M}_{\mathrm{g}}\ge1$. Dust particles are drifted from the outer region of the protoplanetary disk. However, these particles have already grown to the pebbles (cm-sized particles) until they reach around the gas planets like Jupiter \citep[e.g.][]{lam12,oku12,sat16}, so that most of them should be dammed at the outer edge of the gas gap by the positive gas pressure gradient \citep[e.g.][]{ada76,zhu12}. In this case, only a small part of the dust particles can penetrate into the gas gap and flow into the circumplanetary disk, so that the dust-to-gas mass inflow flux ratio should be smaller than unity.

One possibility to achieve the high ratio is considering satellitesimal formation in the final phase of planetary formation. Photoevaporation may increase the dust-to-gas mass ratio in protoplanetary disk as time passes \citep[e.g.][]{ale06a,ale06b}. It is also considered that the gas flux decreases in the final phase because the gas gap becomes wider and deeper \citep[e.g.][]{kan15,tan16}. Our results actually suggested that low gas inflow mass flux is suitable for satellitesimal formation. The midplane temperature $T$ is almost proportional to $\dot{M}_{\mathrm{g}}^{1/4}$ (Equation (\ref{T4})). When the gas inflow decrease, the disk becomes cooler and the snow line moves inward (see Figure \ref{fig:Gas}). This means that the area where icy satellitesimals can form expands. Moreover, the collisional velocity driven by turbulence weakly depends on the gas inflow rate, $\Delta v_{\rm t}\propto c_{\rm s}\propto T^{1/2}\propto\dot{M}_{\mathrm{g}}^{1/8}$ (Equation (\ref{vt})). Low gas inflow mass flux may also contribute to overcoming fragmentation barrier.

\subsection{Effects of the Internal Density} \label{internal}
\begin{figure*}
\epsscale{1.15}
\plotone{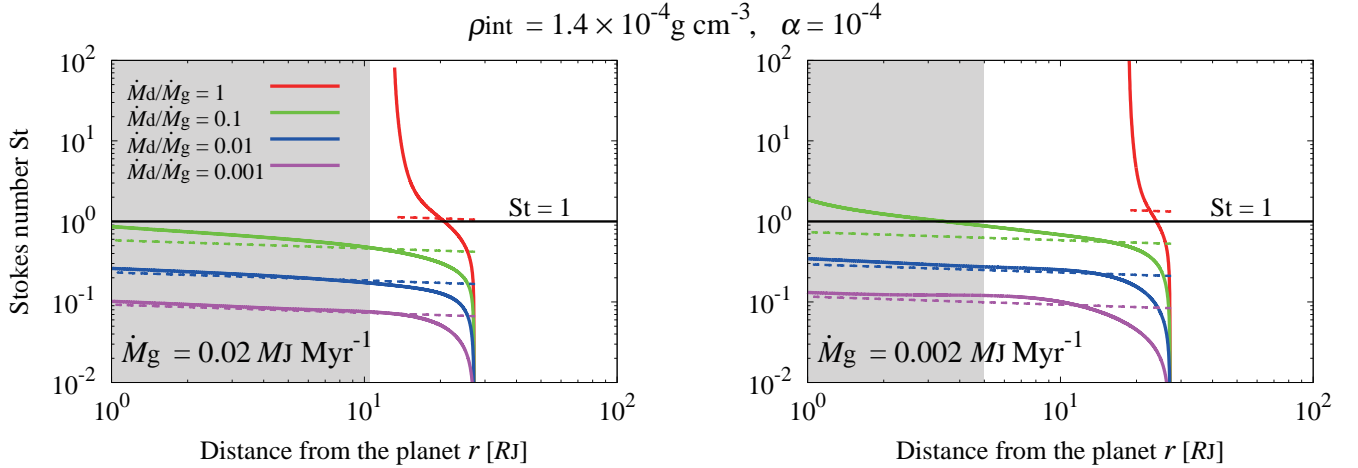}
\caption{Steady-state profiles of the Stokes number ${\rm St}$ of the highly porous ($\rho_{\mathrm{int}}=1.4\times10^{-4} \mathrm{g/cm^{3}}$) dust particles with $\alpha=10^{-4}$, and $\dot{M}_{\mathrm{g}}=0.02$ (left panel) and $0.002~M_{\rm J}~{\rm Myr}^{-1}$ (right panel). The dashed lines in both the panels show the predictions from the analytic estimates given by Equation (\ref{St}). We stopped the calculation when the particle radius reaches $100\ \mathrm{km}$ (the red curve in the left panel). \label{fig:internaldensity}}
\end{figure*}

We investigated the impact of changing the internal density of dust particles on satellitesimal formation. Figure \ref{fig:internaldensity} represents the Stokes number for $\rho_{\mathrm{int}}=1.4\times10^{-4}~\mathrm{g~cm}^{-3}$ and $\alpha=10^{-4}$. We found that the conditions for satellitesimal formation did not change from those with $\rho_{\mathrm{int}}=1.4~\mathrm{g~cm}^{-3}$ (see Figures. \ref{fig:fiducial} and \ref{fig:alpha}). This is because the Stokes number for small $r$ can also be approximated as Equation (\ref{St}) in this case, (we found that $\mathrm{Re_{p}}\ga10^{3}$ for $r\sim10~R_{\mathrm{J}}$ and we have been able to assume $C_{\mathrm{D}}\approx0.5$) and this approximated Stokes number dose not depend on $\rho_{\mathrm{int}}$. In generally, the growth timescale takes a minimum value within the Newton regime \citep{oku12}. Therefore, the Stokes number dose not grow beyond the dashed lines in Figure \ref{fig:internaldensity} by changing $\rho_{\mathrm{int}}$ unless it reaches unity. The impact of changing $\rho_{\mathrm{int}}$ is only that the fluffy particles move earlier than the compact particles from the Stokes regime ($\mathrm{Re_{p}}\la1$) to the Newton regime.

We note that the growth timescale (i.e. $\mathrm{St}$) depends on the internal density only in the Stokes regime \citep{oku12}. Since the gas densities of protoplanetary disks are generally much lower than those of circumplanetary disks, the growth timescale of highly porous dust aggregates in protoplanetary disks can be so small that they overcome the drift barrier within the Stokes regime \citep{oku12}.

\subsection{Streaming Instability} \label{si}
\begin{figure}
\epsscale{1.15}
\plotone{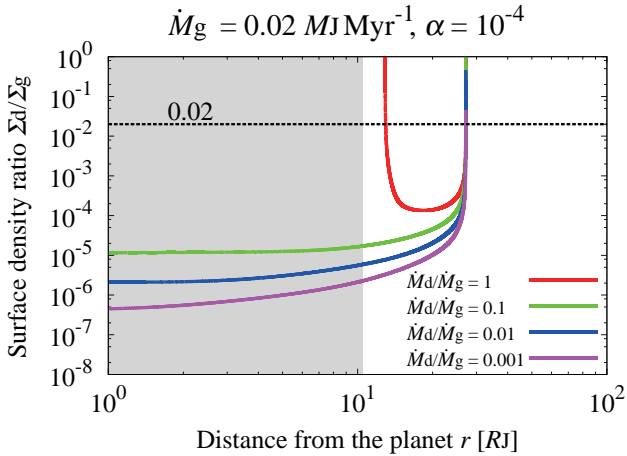}
\caption{Steady-state profiles of the dust-to-gas surface density ratio $\Sigma_{\rm d}/\Sigma_{\rm g}$ with $\dot{M}_{\mathrm{g}}=0.02$ and $\alpha=10^{-4}$. The dashed line shows the occurrence condition of streaming instability in the most suitable case of ${\rm St}$ \citep{car15}. \label{fig:SI}}
\end{figure}

Generating growing particle-density perturbations by streaming instability is another planetesimal formation mechanism not the collisional growth of the dust particles \citep{you05}. The difference between the velocities of the dust and gas drives the instability. The dust particles are concentrated quickly in localized dense clumps, so that they do not drift to the Sun. This mechanism may also be applicable to satellitesimal formation. \citet{car15} showed that the condition that streaming instability is active depends on the dust-to-gas surface density ratio and the Stokes number of the dust particles. They found that the particle clumps can form if $\Sigma_{\mathrm{d}}/\Sigma_{\mathrm{g}}>0.02$ in the most suitable Stokes number condition (see Figure 8 in \citet{car15}). However, our results showed that the ratio is much lower than the critical value (see Figure \ref{fig:SI}), so that it should be difficult to form satellitesimals via the streaming instability process. Note that water sublimation or recondensation near the snowline can result in an enhancement in the dust surface density and it may be able to trigger streaming instability \citep{ida16a,ida16b,sch17}.

\subsection{Validity of the Single-size Approach in Circumplanetary Disks} \label{validity}
We used a single-size approach to investigate the growth and drift of dust particles in circumplanetary disks. As shown by \citet{sat16}, this approach is valid for the growth and drift of mass-dominating particles in protoplanetary disks, as long as the collisional destruction of the particles is negligible (see also \citet{kri16}, \citet{oku16}, and \citet{tsu17} for applications to dust growth in protoplanetary disks).

In principle, the single-size approximation breaks down when there is more than one population of particles that dominates the total dust mass. Unlike our assumption that the dust inflow onto the circumplanetary disk is concentrated at the outer edge of the gas inflow region, the dust inflow may be extended over a wide area of the disk. In this case, the size distribution in the inner disk regions may have two peaks of the drifting pebbles that accreted in the outer disk region and the small dust grains directly supplied to the inner disk regions. Actually, when the dust particles strongly couple with the gas, the mass flux of the dust inflow should be proportional to that of the gas $f\propto r^{-1}$. The ratio of the dust mass flux flowing to the inside of $r$ relative to the total dust inflow mass flux is then $\approx r/r_{\mathrm{b}}$ which is still about $0.5$ even if $r=15~R_{\rm J}$ (the orbit of Ganymede). Therefore, the actual size distribution of the dust particles may be wide. We note that our assumption of the concentrated dust inflow may have also caused overestimation of the Stoke number especially in the outer disk regions ($\mathrm{St}\propto \dot{M}_{\mathrm{d}}^{2/5}$, $\dot{M}_{\mathrm{d}}^{2/3}$, or $\dot{M}_{\mathrm{d}}^{6/11}$ in our model, see Equations (\ref{St}), (\ref{Stg}), and (\ref{Stp})) and the satellitesimal formation could be harder in reality.

On the other hand, even if the dust particles have a wide size distribution, the smaller particles should grow rapidly and the distribution will narrow. This is because the dust growth timescale $t_{\rm grow}$ must be an increasing function of the dust mass $m_{\rm d}$  in the Newton regime ($t_{\rm grow}\propto m_{\rm d}^{1/4}$ or $m_{\rm d}^{1/3}$ in our model, see Equations (\ref{growth}), (\ref{continuous}), (\ref{stokes_newton}), and the approximations in Section \ref{gasdust} and Appendix \ref{app}) so that the dust particles should grow as orderly growth in most areas within the disk. However, the $m_{\rm d}$ dependence of $t_{\rm grow}$ is weak and we will have to make sure of the validity of the single-size approach by using full-size calculations of dust growth in circumplanetery disks in future.

\section{Conclusions}
\label{conclusions}
We have investigated whether icy dust particles can form satellitesimals by their pairwise collisional growth in circumplanetary disks with various conditions. We have calculated the distributions of the surface density, radius and Stokes number of the peak mass dust particles. Our model considered only steady conditions and assumed that the temperature of the circumplanetary disk is almost like a minimum estimate. The gas inflow and the circumplanetary disk model were based on the results of \citet{tan12} and \citet{fuj14}. We have changed the dust and gas inflow mass fluxes and the strength of turbulence in the disk so that we understood the effects of these factors to satellitesimal formation. We have also approximated the Stokes number (normalized stopping time) of dust particles to understand these effects because satellitesimals can form only when the Stokes number achieves unity. From these parameter studies, we have revealed the conditions for satellitesimal formation and discussed the feasibilities of them. We have also discussed the effects of changing the internal density of dust particles and the possibility of satellitesimal formation by streaming instability.  Our findings are summarized as follows.

\begin{enumerate}
\item The mass flux of dust and gas flowing into the circumplanetary disk determines the evolution of dust particles. Especially, the dust-to-gas inflow mass flux ratio, $\dot{M}_{\mathrm{d}}/\dot{M}_{\mathrm{g}}$, is important. A larger dust inflow provides a higher dust density in the circumplanetary disk, accelerating collisional dust growth. A small amount of gas inflow makes the gas surface density of the circumplanetary disk lower and the gas drag force that dust particles receive weaker. This reduces the radial drift speed of the particles and this is advantageous for  collisional growth. The approximated Stokes number is actually proportional to $(\dot{M}_{\mathrm{d}}/\dot{M}_{\mathrm{g}})^{2/5}$, $(\dot{M}_{\mathrm{d}}/\dot{M}_{\mathrm{g}})^{2/3}$, or $(\dot{M}_{\mathrm{d}}/\dot{M}_{\mathrm{g}})^{6/11}$ (Equations (\ref{St}), (\ref{Stg}), and (\ref{Stp})). As a result, satellitesimals can form only for $\dot{M}_{\mathrm{d}}/\dot{M}_{\mathrm{g}}\ge1$. For fixed $\dot{M}_{\mathrm{d}}/\dot{M}_{\mathrm{g}}$, the Stokes number depends on the gas inflow mass flux only $\dot{M}_{\mathrm{g}}^{1/10}$, $\dot{M}_{\mathrm{g}}^{7/36}$, or $\dot{M}_{\mathrm{g}}^{5/44}$ (Equations (\ref{St}), (\ref{Stg}), and (\ref{Stp})), so that the key parameter is not the pure gas (or pure dust) inflow mass flux but the ratio of the dust and gas muss inflow.
\item The strength of turbulence also affects the fate of dust particles. When $\dot{M}_{\mathrm{g}}=0.02~M_{\rm J}~{\rm Myr}^{-1}$, the approximated Stokes number is proportional to $\alpha^{1/5}$ or $\alpha^{1/11}$ (Equations (\ref{St}) and (\ref{Stp})). As a result, satellitesimals can form only for $\alpha\ge10^{-4}$. Moreover, strong turbulence increases the collisional velocity and it actually exceeds the critical fragmentation velocity of $50\ \mathrm{m~s^{-1}}$ \citep{wad09} when $\alpha=10^{-2}$ (Figure \ref{fig:velocities}).
\end{enumerate}
In summary, the conditions for satellitesimal formation via collisional dust growth are $\dot{M}_{\mathrm{d}}/\dot{M}_{\mathrm{g}}\ge1$ and $10^{-4}\le\alpha<10^{-2}$ when $\dot{M}_{\mathrm{g}}=0.02~M_{\rm J}~{\rm Myr}^{-1}$ (Figure \ref{fig:condition}). This result does not strongly depend on the gas inflow mass flux.

Our results also suggest that the porosity of dust particles does not affect the condition for satellitesimal formation (Section \ref{internal}). The Stokes number of dust particles take the largest value in the Newton regime and the value does not depend on the internal density (Equation (\ref{St})). Low internal density just makes the particles go into the Newton regime quickly.

In reality, it would be difficult to achieve the condition $\dot{M}_{\mathrm{d}}/\dot{M}_{\mathrm{g}}\ge1$ (Section \ref{feasibility}). In protoplanetary disks, the majority of dust in the vicinity of gas giants should settle down to the midplane so that the inflow gas from the high altitude may be dust-poor. The dust particles also have already grown to pebble-sized particles. Such large particles may not accrete onto circumplanetary disks, because they are easily trapped at the edges of planet-carved gas gaps \citep{zhu12}.

Satellitesimal formation via streaming instability is also unfeasible (Section \ref{si}). The dust-to-gas surface density ratio $\Sigma_{\mathrm{d}}/\Sigma_{\mathrm{g}}$ is much smaller than the critical value 0.02 that the instability occurs.

However, the photoevaporation may contribute to satisfy the condition for satellitesimal formation by increasing the dust-to-gas ratio of the inflow gas (Section \ref{feasibility}). The final phase of planetary formation should be also suitable for satellitesimal formation because of the wide and deep gas gap. The smaller gas inflow mass flux becomes, the lower temperature and the slower collision velocity become.

The delivery of planetesimals from the protoplanetary disk may be necessary for satellite formation around gas giants. As we mentioned in Section \ref{introduction}, the planetesimal capture hypothesis can roughly reproduce the initial radial distribution of satellitesimals assumed in the previous satellite formation models \citep{sue17}. Moreover, the captured planetesimals may efficiently accrete the drifting dust particles which could not grow to satellitesimals in the circumplanetary disk. Such a satellite formation scenario via pebble accretion will be investigated in our future work.

\acknowledgments
We thank the anonymous reviewer for very useful comments. We also thank Yann Alibert, Sebastien Charnoz, Joanna Drazkowska, Cornelis Dullemond, Paul Estrada, Yuri Fujii, Ryuki Hyodo, Yamila Miguel, Chris Ormel, Judit Szul\'{a}gyi, and Takayuki Tanigawa for valuable discussions. This work was supported by JSPS KAKENHI Grant Numbers JP15H02065, JP16K17661, JP16H04081, JP16J09590.

\appendix
\section{Approximation of the Stokes Number for Low Gas Density Cases} \label{app}
Although we assume that $C_{\rm D}$ is a constant in Section \ref{gasdust}, this assumption is not correct when the gas density is low by small gas inflow mass flux or strong turbulence. When $\dot{M}_{\rm{g}}=0.002~M_{\rm J}~{\rm Myr}^{-1}$ and $\alpha=10^{-4}$, or $\dot{M}_{\rm{g}}=0.002~M_{\rm J}~{\rm Myr}^{-1}$ and $\alpha=10^{-2}$, the particle Reynolds number is with in the range of $10^{-1}\la{\rm Re_{p}}\la10^{2}$ for $r\sim10~R_{\rm J}$. In this case, $C_{\rm D}$ can be approximated as $C_{\rm{D}}\approx12/\sqrt{\rm{Re_{p}}}$ (Equation (\ref{CD})). When $\dot{M}_{\rm{g}}=0.002~M_{\rm J}~{\rm Myr}^{-1}$ and $\alpha=10^{-4}$, the dust--dust and dust--gas relative velocities are approximated as $\Delta v_{\rm{dd}}\approx(1/2)v_{\rm{r}}$ and $\Delta v_{\rm{dg}}\approx v_{\rm{r}}$. The Stokes number $\rm{St}$ can then be approximated as
\begin{equation}
\mathrm{St}\approx1.6\left(\dfrac{\dot{M}_{\mathrm{d}}/\dot{M}_{\mathrm{g}}}{1}\right)^{2/3}\left(\dfrac{\alpha}{10^{-4}}\right)^{-1/9}\left(\dfrac{\dot{M}_{\mathrm{g}}}{0.002~M_{\mathrm{J}}~{\mathrm{Myr}}^{-1}}\right)^{4/9}\left(\dfrac{T}{90~\rm{K}}\right)^{-1}
\left(\dfrac{\rho_{\mathrm{int}}}{1.4~\mathrm{g/cm^{3}}}\right)^{-2/9}\left(\dfrac{M_{\mathrm{cp}}}{1~M_{\mathrm{J}}}\right)^{7/9}\left(\dfrac{r}{10~R_{\mathrm{J}}}\right)^{-13/9},
\label{Stg}
\end{equation}
for small $r$. Unlike in Equations (\ref{St}) and (\ref{Stp}), $\mathrm{St}$ decreases with increasing $\alpha$, although the dependence is very weak. Strong turbulence diffuses dust particles into the vertical direction and thereby reduces their collision rate (see Equations (\ref{growth}) and (\ref{Hd})). However, this effect is canceled out by the particle collision velocity induced by turbulence, which increases with increasing $\alpha$. When $\dot{M}_{\rm{g}}=0.02~M_{\rm J}~{\rm Myr}^{-1}$ and $\alpha=10^{-2}$, the two relative velocities are determined by the strength of turbulence (i.e. $\Delta v_{\rm{dd}}\approx\Delta v_{\rm{dg}}\approx\Delta v_{\rm{t}}\approx\sqrt{3\alpha}c_{\mathrm{s}}\mathrm{St}_{1}^{1/2}$, see Figure \ref{fig:velocities}),
\begin{equation}
\mathrm{St}\approx0.73\left(\dfrac{\dot{M}_{\mathrm{d}}/\dot{M}_{\mathrm{g}}}{0.1}\right)^{6/11}\left(\dfrac{\alpha}{10^{-2}}\right)^{1/11}\left(\dfrac{\dot{M}_{\mathrm{g}}}{0.02~M_{\rm J}~{\rm Myr}^{-1}}\right)^{4/11}\left(\dfrac{T}{160~\rm{K}}\right)^{-1}\left(\dfrac{\rho_{\mathrm{int}}}{1.4~\mathrm{g/cm^{3}}}\right)^{-2/11}\left(\dfrac{M_{\mathrm{cp}}}{1~M_{\mathrm{J}}}\right)^{9/11}\left(\dfrac{r}{10~R_{\mathrm{J}}}\right)^{-15/11},
\label{Stp}
\end{equation}
for small $r$.

\bibliography{bib2}

\begin{thebibliography}{}
\expandafter\ifx\csname natexlab\endcsname\relax\def\natexlab#1{#1}\fi

\bibitem[{Adachi {et~al.}(1976)Adachi, Hayashi, \& Nakazawa}]{ada76}
Adachi, I., Hayashi, C., \& Nakazawa, K. 1976, Progress of Theoretical Physics,
  56, 1756

\bibitem[{Alexander {et~al.}(2006{\natexlab{a}})Alexander, Clarke, \&
  Pringle}]{ale06a}
Alexander, R.~D., Clarke, C.~J., \& Pringle, J.~E. 2006{\natexlab{a}}, Monthly
  Notices of the Royal Astronomical Society, 369, 216

\bibitem[{Alexander {et~al.}(2006{\natexlab{b}})Alexander, Clarke, \&
  Pringle}]{ale06b}
---. 2006{\natexlab{b}}, Monthly Notices of the Royal Astronomical Society,
  369, 229

\bibitem[{Alibert {et~al.}(2005)Alibert, Mousis, \& Benz}]{ali05}
Alibert, Y., Mousis, O., \& Benz, W. 2005, Astronomy \& Astrophysics, 439, 1205

\bibitem[{Blum \& Wurm(2008)}]{blu08}
Blum, J., \& Wurm, G. 2008, Annu. Rev. Astron. Astrophys., 46, 21

\bibitem[{Canup \& Ward(2002)}]{can02}
Canup, R.~M., \& Ward, W.~R. 2002, The Astronomical Journal, 124, 3404

\bibitem[{Canup \& Ward(2006)}]{can06}
---. 2006, Nature, 441, 834

\bibitem[{Carrera {et~al.}(2015)Carrera, Johansen, \& Davies}]{car15}
Carrera, D., Johansen, A., \& Davies, M.~B. 2015, Astronomy \& Astrophysics,
  579, A43

\bibitem[{{D'Angelo} {et~al.}(2002){D'Angelo}, {Henning}, \& {Kley}}]{dan02}
{D'Angelo}, G., {Henning}, T., \& {Kley}, W. 2002, \aap, 385, 647

\bibitem[{D'Angelo \& Podolak(2015)}]{dan15}
D'Angelo, G., \& Podolak, M. 2015, The Astrophysical Journal, 806, 203

\bibitem[{Estrada {et~al.}(2009)Estrada, Mosqueira, Lissauer, D'Angelo, \&
  Cruikshank}]{est09}
Estrada, P.~R., Mosqueira, I., Lissauer, J., D'Angelo, G., \& Cruikshank, D.
  2009, Europa, edited by RT Pappalardo, WB McKinnon, and K. Khurana,
  University of Arizona Press, Tucson, 27

\bibitem[{Fujii {et~al.}(2014)Fujii, Okuzumi, Tanigawa, \& ichiro
  Inutsuka}]{fuj14}
Fujii, Y.~I., Okuzumi, S., Tanigawa, T., \& ichiro Inutsuka, S. 2014, The
  Astrophysical Journal, 785, 101

\bibitem[{Fujita {et~al.}(2013)Fujita, Ohtsuki, Tanigawa, \& Suetsugu}]{fuj13}
Fujita, T., Ohtsuki, K., Tanigawa, T., \& Suetsugu, R. 2013, The Astronomical
  Journal, 146, 140

\bibitem[{Fung \& Chiang(2016)}]{fun16}
Fung, J., \& Chiang, E. 2016, The Astrophysical Journal, 832, 105

\bibitem[{Gundlach \& Blum(2015)}]{gun15}
Gundlach, B., \& Blum, J. 2015, The Astrophysical Journal, 798, 34

\bibitem[{Hayashi(1981)}]{hay81}
Hayashi, C. 1981, Progress of Theoretical Physics Supplement, 70, 35

\bibitem[{Ida \& Guillot(2016)}]{ida16b}
Ida, S., \& Guillot, T. 2016, Astronomy \& Astrophysics, 596, L3

\bibitem[{Ida {et~al.}(2016)Ida, Guillot, \& Morbidelli}]{ida16a}
Ida, S., Guillot, T., \& Morbidelli, A. 2016, Astronomy \& Astrophysics, 591,
  A72

\bibitem[{Kanagawa {et~al.}(2015)Kanagawa, Muto, Tanaka, Tanigawa, Takeuchi,
  Tsukagoshi, \& Momose}]{kan15}
Kanagawa, K.~D., Muto, T., Tanaka, H., {et~al.} 2015, The Astrophysical Journal
  Letters, 806, L15

\bibitem[{Kobayashi {et~al.}(2012)Kobayashi, Ormel, \& Ida}]{kob12}
Kobayashi, H., Ormel, C.~W., \& Ida, S. 2012, The Astrophysical Journal, 756,
  70

\bibitem[{Krijt {et~al.}(2016)Krijt, Ormel, Dominik, \& Tielens}]{kri16}
Krijt, S., Ormel, C.~W., Dominik, C., \& Tielens, A.~G. 2016, Astronomy \&
  Astrophysics, 586, A20

\bibitem[{Lambrechts \& Johansen(2012)}]{lam12}
Lambrechts, M., \& Johansen, A. 2012, Astronomy \& Astrophysics, 544, A32

\bibitem[{Lubow {et~al.}(1999)Lubow, Seibert, \& Artymowicz}]{lub99}
Lubow, S.~H., Seibert, M., \& Artymowicz, P. 1999, The Astrophysical Journal,
  526, 1001

\bibitem[{Lunine \& Stevenson(1982)}]{lun82}
Lunine, J.~I., \& Stevenson, D.~J. 1982, Icarus, 52, 14

\bibitem[{Miguel \& Ida(2016)}]{mig16}
Miguel, Y., \& Ida, S. 2016, Icarus, 266, 1

\bibitem[{Mosqueira {et~al.}(2010)Mosqueira, Estrada, \& Turrini}]{mos10}
Mosqueira, I., Estrada, P., \& Turrini, D. 2010, Space Science Reviews, 153,
  431

\bibitem[{Mosqueira \& Estrada(2003{\natexlab{a}})}]{mos03a}
Mosqueira, I., \& Estrada, P.~R. 2003{\natexlab{a}}, Icarus, 163, 198

\bibitem[{Mosqueira \& Estrada(2003{\natexlab{b}})}]{mos03b}
---. 2003{\natexlab{b}}, Icarus, 163, 232

\bibitem[{{Nakamoto} \& {Nakagawa}(1994)}]{nak94}
{Nakamoto}, T., \& {Nakagawa}, Y. 1994, \apj, 421, 640

\bibitem[{Ogihara \& Ida(2012)}]{ogi12}
Ogihara, M., \& Ida, S. 2012, The Astrophysical Journal, 753, 60

\bibitem[{Okuzumi {et~al.}(2016)Okuzumi, Momose, iti Sirono, Kobayashi, \&
  Tanaka}]{oku16}
Okuzumi, S., Momose, M., iti Sirono, S., Kobayashi, H., \& Tanaka, H. 2016, The
  Astrophysical Journal, 821, 82

\bibitem[{Okuzumi {et~al.}(2012)Okuzumi, Tanaka, Kobayashi, \& Wada}]{oku12}
Okuzumi, S., Tanaka, H., Kobayashi, H., \& Wada, K. 2012, The Astrophysical
  Journal, 752, 106

\bibitem[{Ormel \& Cuzzi(2007)}]{orm07}
Ormel, C., \& Cuzzi, J. 2007, Astronomy \& Astrophysics, 466, 413

\bibitem[{Perets \& Murray-Clay(2011)}]{per11}
Perets, H.~B., \& Murray-Clay, R.~A. 2011, The Astrophysical Journal, 733, 56

\bibitem[{Ros \& Johansen(2013)}]{ros13}
Ros, K., \& Johansen, A. 2013, Astronomy \& Astrophysics, 552, A137

\bibitem[{Saito \& Sirono(2011)}]{sai11}
Saito, E., \& Sirono, S. 2011, The Astrophysical Journal, 728, 20

\bibitem[{Sasaki {et~al.}(2010)Sasaki, Stewart, \& Ida}]{sas10}
Sasaki, T., Stewart, G.~R., \& Ida, S. 2010, The Astrophysical Journal, 714,
  1052

\bibitem[{Sato {et~al.}(2016)Sato, Okuzumi, \& Ida}]{sat16}
Sato, T., Okuzumi, S., \& Ida, S. 2016, Astronomy \& Astrophysics, 589, A15

\bibitem[{Schoonenberg \& Ormel(2017)}]{sch17}
Schoonenberg, D., \& Ormel, C.~W. 2017, Astronomy \& Astrophysics

\bibitem[{{Shakura} \& {Sunyaev}(1973)}]{sha73}
{Shakura}, N.~I., \& {Sunyaev}, R.~A. 1973, \aap, 24, 337

\bibitem[{Sohl {et~al.}(2002)Sohl, Spohn, Breuer, \& Nagel}]{soh02}
Sohl, F., Spohn, T., Breuer, D., \& Nagel, K. 2002, Icarus, 157, 104

\bibitem[{Suetsugu \& Ohtsuki(2017)}]{sue17}
Suetsugu, R., \& Ohtsuki, K. 2017, The Astrophysical Journal, 839, 66

\bibitem[{Suetsugu {et~al.}(2016)Suetsugu, Ohtsuki, \& Fujita}]{sue16}
Suetsugu, R., Ohtsuki, K., \& Fujita, T. 2016, The Astronomical Journal, 151,
  140

\bibitem[{Szul{\'a}gyi {et~al.}(2016)Szul{\'a}gyi, Masset, Lega, Crida,
  Morbidelli, \& Guillot}]{szu16a}
Szul{\'a}gyi, J., Masset, F., Lega, E., {et~al.} 2016, Monthly Notices of the
  Royal Astronomical Society, 460, 2853

\bibitem[{Tanigawa {et~al.}(2014)Tanigawa, Maruta, \& Machida}]{tan14}
Tanigawa, T., Maruta, A., \& Machida, M.~N. 2014, The Astrophysical Journal,
  784, 109

\bibitem[{Tanigawa {et~al.}(2012)Tanigawa, Ohtsuki, \& Machida}]{tan12}
Tanigawa, T., Ohtsuki, K., \& Machida, M.~N. 2012, The Astrophysical Journal,
  747, 47

\bibitem[{Tanigawa \& Tanaka(2016)}]{tan16}
Tanigawa, T., \& Tanaka, H. 2016, The Astrophysical Journal, 823, 48

\bibitem[{Tsukamoto {et~al.}(2017)Tsukamoto, Okuzumi, \& Kataoka}]{tsu17}
Tsukamoto, Y., Okuzumi, S., \& Kataoka, A. 2017, The Astrophysical Journal,
  838, 151

\bibitem[{Turner {et~al.}(2014)Turner, Lee, \& Sano}]{tur14}
Turner, N.~J., Lee, M.~H., \& Sano, T. 2014, The Astrophysical Journal, 783, 14

\bibitem[{Wada {et~al.}(2009)Wada, Tanaka, Suyama, Kimura, \& Yamamoto}]{wad09}
Wada, K., Tanaka, H., Suyama, T., Kimura, H., \& Yamamoto, T. 2009, The
  Astrophysical Journal, 702, 1490

\bibitem[{Ward \& Canup(2010)}]{war10}
Ward, W.~R., \& Canup, R.~M. 2010, The Astronomical Journal, 140, 1168

\bibitem[{Weidenschilling(1977)}]{wei77}
Weidenschilling, S. 1977, Icarus, 44, 172

\bibitem[{Whipple(1972)}]{whi72}
Whipple, F. 1972, From plasma to planet, ed. A. Elvius (London: Wiley)

\bibitem[{Youdin \& Goodman(2005)}]{you05}
Youdin, A.~N., \& Goodman, J. 2005, The Astrophysical Journal, 620, 459

\bibitem[{Youdin \& Lithwick(2007)}]{you07}
Youdin, A.~N., \& Lithwick, Y. 2007, Icarus, 192, 588

\bibitem[{Zhu {et~al.}(2012)Zhu, Nelson, Dong, Espaillat, \& Hartmann}]{zhu12}
Zhu, Z., Nelson, R.~P., Dong, R., Espaillat, C., \& Hartmann, L. 2012, The
  Astrophysical Journal, 755, 6

\end{thebibliography}

\end{document}